\documentclass[sigconf,nonacm]{acmart}

\usepackage{multirow}
\usepackage{caption}
\usepackage{colortbl}
\usepackage{graphicx}
\usepackage{siunitx}

\AtBeginDocument{%
  }

\begin{document}

\title{Benchmarking for Single Feature Attribution with Microarchitecture Cliffs}

\author{Hao Zhen}
\authornote{Hao Zhen and Qingxuan Kang contributed equally to this work.}
\affiliation{
  \institution{State Key Lab of Processors, Institute of Computing Technology, Chinese Academy of Sciences}
  \country{China}
}

\author{Qingxuan Kang}
\authornotemark[1]
\affiliation{
  \institution{National University of Singapore}
  \country{Singapore}
}

\author{Yungang Bao}
\affiliation{
  \institution{State Key Lab of Processors, Institute of Computing Technology, Chinese Academy of Sciences}
  \country{China}
}

\author{Trevor E. Carlson}
\affiliation{
  \institution{National University of Singapore}
  \country{Singapore}
}

\begin{abstract}

Architectural simulators play a critical role in early microarchitectural exploration due to their flexibility and high productivity.
However, their effectiveness is often constrained by fidelity: simulators may deviate from the behavior of the final RTL, leading to unreliable performance estimates.
Consequently, model calibration, which aligns simulator behavior with the RTL as the ground-truth microarchitecture, becomes essential for achieving accurate performance modeling.

To facilitate model calibration accuracy, we propose \textit{Microarchitecture Cliffs}, a benchmark generation methodology designed to expose mismatches in microarchitectural behavior between the simulator and RTL.
After identifying the key architectural components that require calibration,
the Cliff methodology enables precise attribution of microarchitectural differences to a single microarchitectural feature through a set of benchmarks. 
In addition, we develop a set of automated tools to improve the efficiency of the Cliff workflow.

We apply the Cliff methodology to calibrate the XiangShan version of gem5 (XS-GEM5) against the XiangShan open-source CPU (XS-RTL). 
We reduce the performance error of XS-GEM5 from 59.2\% to just 1.4\% on the Cliff benchmarks.
Meanwhile, the calibration guided by Cliffs effectively reduces the relative error of a representative tightly coupled microarchitectural feature by 48.03\%.
It also substantially lowers the absolute performance error, with reductions of 15.1\% and 21.0\% on SPECint2017 and SPECfp2017, respectively.
\end{abstract}


\maketitle

\section{Introduction} \label{introduction}

Architectural simulators are widely used for performance exploration during chip development due to their flexibility for design modification, fast build times, and efficient execution~\cite{binkert2011gem5,oh2024udp}.
In practice, new microarchitectural features are typically implemented and evaluated on simulators first, then validated on the RTL design.
This two-stage workflow significantly improves the efficiency of exploration.

However, this workflow faces a fundamental fidelity challenge~\cite{desikan2001measuring,nowatzki2015architectural,black1998calibration}: simulators may not fully capture the behavior of the final RTL.
Such discrepancies can lead to inaccurate performance estimates and undermine the reliability of early-stage design exploration.
In particular, because of their higher abstraction level, simulators are more prone to modeling mismatches caused by outdated parameter configurations, idealized or incomplete resource-contention modeling, or misestimated microarchitectural feature importance~\cite{gutierrez2014sources}.

Model calibration~\cite{pathak2025towards,huppert2021memory,pimentel2008calibration}, which ensures that the simulator consistently reflects the intended microarchitecture, thus becomes essential.
It leverages authoritative design knowledge, such as detailed specifications, feature lists, and RTL source code, to identify modeling gaps, validate assumptions, and refine simulator configurations so that its behavior matches the RTL, the ground-truth microarchitectural intent.
A natural first step is \textit{specification-driven calibration}, where key modeling objectives are extracted directly from design specification and RTL.
However, as modern architectures grow more complex, manually tuning numerous configuration knobs becomes increasingly impractical and error-prone~\cite{zhang2024dataflow,qiu2023gem5tune}.
Moreover, the performance impact of certain design features is not always obvious.
Simulators may simplify them for flexibility~\cite{sanchez2013zsim,renda2020difftune}, only to later find that these details significantly affect performance in specific scenarios.

\begin{figure}[tb]
\centering
\includegraphics[width=0.48\textwidth]{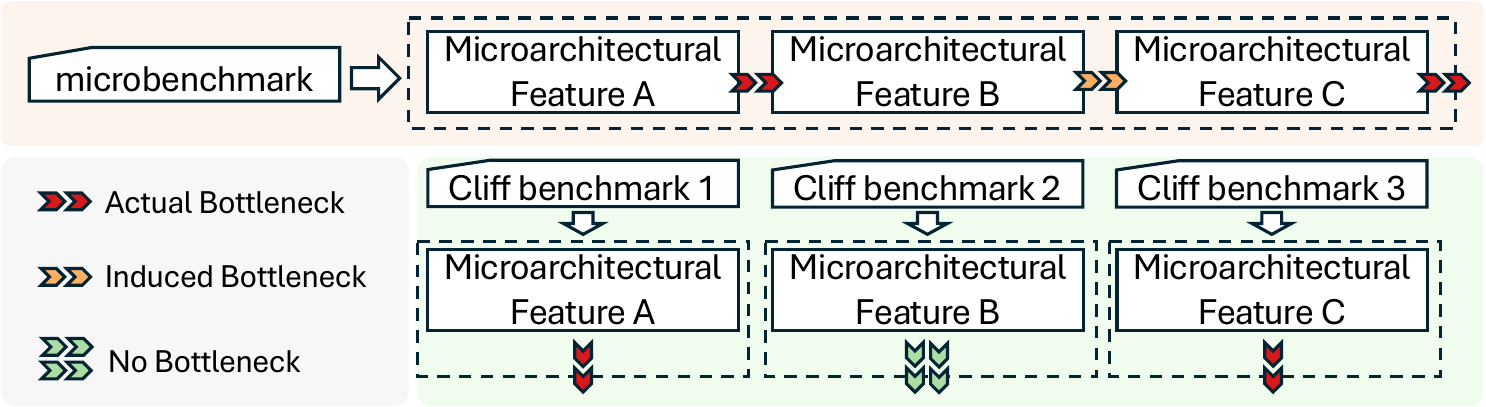}
\caption{Microbenchmark vs. Cliffs calibration. Microbenchmark confounds the effect of multiple microarchitectural features, while Cliff benchmarks isolate each feature individually.}
\label{fig:cliff-illust}
\end{figure}

Given these limitations, \textit{behavior-driven calibration} provides a complementary perspective by examining software-visible performance behaviors (e.g., IPC) to expose mismatches between the simulator and the RTL.
A common technique is microbenchmarking, which uses small, synthetic code kernels to probe specific microarchitectural features~\cite{yotov2005x}.
However, microbenchmarks often inadvertently exercise multiple interacting microarchitectural components, making it difficult to isolate the contribution of any single microarchitectural feature~\cite{nasr2025concorde}.
For example, a microbenchmark intended to measure DCache latency is found to be confounded by unrelated microarchitectural behaviors, such as load-violation checking and recovery penalties from load-use mis-speculation~\cite{desikan2001measuring}.

To facilitate efficient and unambiguous calibration,
we propose \textit{Microarchitecture Cliffs (Cliffs)}\footnote{We will open source the tools upon acceptance.},  
a systematic testing methodology that attributes discrepancies to a single microarchitectural feature, thereby enabling an effective assessment of microarchitectural-level calibration accuracy.
Each Cliff benchmark targets a specific microarchitectural feature, 
such as a distinct allocation policy (e.g., ROB allocation) or a resource contention constraint (e.g., DCache bank conflicts).

Microbenchmarking and Cliffs differ in granularity.
As shown in Fig.~\ref{fig:cliff-illust}, a microbenchmark may simultaneously activate several microarchitectural features.
When one or more performance bottlenecks are present (\textit{red} arrows), identifying all of them may not be straightforward, as it is difficult to distinguish them from those that are merely \textit{induced} by upstream constraints (\textit{yellow} arrows, feature B).
In contrast, Cliffs isolate individual microarchitectural features, surface bottlenecks without interference from other components, and correctly identify true bottlenecks (features A and C).

To realize the Cliff methodology, we address two key challenges.
(1) \textit{Given the large state space of modern microarchitectures, how can we identify which components require calibration?}
Cliffs cluster performance counters that reveal performance bottlenecks and use them to prioritize microarchitectural components for calibration.
(2) \textit{Given the complex dependencies among microarchitectural components, how can we construct Cliff benchmarks that enable single-feature attribution?}
Guided by the prioritized components and feature list, the Cliff methodology constructs instruction snippets that progressively stress and isolate a single microarchitectural feature, allowing reliable single-feature attribution.

Using the Cliff methodology,
we curate a set of Cliff benchmarks to support the calibration between the XiangShan version of gem5 (XS-GEM5) and XiangShan open-source CPU (XS-RTL)~\cite{xu2022towards,xsgem5_github,xsrtl_github,xiangshan}.
During calibration, the performance error between XS-GEM5 and XS-RTL across Cliffs benchmarks is reduced from 59.2\% to just 1.4\%.
We further demonstrate that, through detailed microarchitectural calibration, the Cliff methodology significantly improves the accuracy of the measured performance gain of the evaluated feature (\textit{relative error}).
More specifically, we use Store Set, a memory dependency predictor, as a representative \textit{tightly coupled microarchitectural feature} because its effectiveness depends on several interacting microarchitectural mechanisms, including \textit{Nuke Replay} and \textit{STA-STD separation} (described in $\S$\ref{storeset}).
By isolating and calibrating these coupled features, Cliffs reduces the relative error of the Store Set predictor’s evaluation on \textit{h264ref} by 48.03\%.
Meanwhile, Cliffs also effectively reduce the performance discrepancy between the simulator and RTL (\textit{absolute performance error}).
The impact of this calibration is demonstrated by reduced runtime differences:
7.6\% (\textit{int}), 15.1\% (\textit{fp}) for SPEC CPU2006 and 15.1\% (\textit{int}), 21.0\% (\textit{fp}) for SPEC CPU2017.

In summary, we make the following major contributions:
\begin{itemize}
    \item We find that isolating performance differences to a single feature improves simulator-RTL calibration.
    \item We propose Microarchitecture Cliffs, a benchmark generation methodology that attributes performance discrepancies between the architectural simulator and RTL to a single microarchitectural feature.
    \item Guided by Cliffs, we systematically calibrate XS-GEM5 against XS-RTL, reducing the relative error when using the Store Set predictor by 48.03\%.
    \item Guided by Cliffs, we also significantly reduced the absolute performance error between XS-GEM5 and XS-RTL.
\end{itemize}

\section{Background} \label{background}

To accurately project performance trends for a specific microarchitectural design, the architectural simulator should be carefully calibrated to reflect the corresponding design features and constraints.
Unlike reverse engineering~\cite{ren2021see,nam2024dramscope,liu2024uncovering,liu2025mdpeek}, calibration treats the microarchitecture as a white box.
Thus, a systematic calibration process typically involves two complementary approaches: \textit{specification-driven calibration} ($\S$\ref{b-forwardanalysis}) and \textit{behavior-driven calibration} ($\S$\ref{background_validation}).

\subsection{Specification-Driven Calibration}
\label{b-forwardanalysis}

Specification-driven calibration begins with architects identifying performance-critical features from the design specification and examining their possible mismatches in the existing simulation infrastructure.
Once such mismatches are identified as likely sources of performance inaccuracy, the modeling team begins implementing the corresponding feature to capture the relevant microarchitectural details.
However, as hardware designs grow increasingly complex to support diverse microarchitectural features, achieving comprehensive calibration through first principles becomes highly challenging.
Identifying performance-critical insights from a microarchitectural design specification requires enumerating the possible interactions among multiple components and reasoning about the concurrency, contention, and feedback loops that determine the efficiency of ILP and MLP~\cite{hammond2004transactional,moore2006logtm,carlson2011sniper}.
Accurately modeling these features, including speculative mechanisms and concurrent pipeline events, requires substantial domain expertise and modeling effort.
The structural and behavioral complexity of these components makes critical microarchitectural details prone to modeling mismatches~\cite{nowatzki2015architectural}.
As a result, \textit{behavior-driven calibration} is required to verify whether the modeled behavior meets expectations by observing the resulting performance characteristics.

\subsection{Behavior-Driven Calibration}
 \label{background_validation}
Behavior-driven calibration identifies modeling mismatches by using targeted benchmarks and comparing performance metrics, such as IPC or microarchitectural events, between the simulator and the RTL.
This process is analogous to RTL design and verification: implementing RTL from a design specification is a specification-driven calibration process.
However, bugs in RTL implementations require systematic verification using
test cases designed to probe corner cases~\cite{constantinides2008online,vasudevan2021learning}. 
Similarly, in performance modeling and calibration, developing an architectural simulator based on RTL code and feature lists is a specification-driven calibration process, and behavior-driven calibration is crucial for identifying implementation discrepancies between the simulator and the RTL. 
Therefore, performance convergence using targeted behavior-driven calibration approaches is as essential to architectural simulators as functional verification is to RTL development.
Under the behavior-driven calibration approach, we observe three methods are commonly used: \textit{Microbenchmarking}, \textit{Time-Proportional}, and \textit{Performance Event Analysis}.

\textbf{Microbenchmarking:}
Microbenchmarks use instruction snippets to analyze specific operations, 
enabling fast evaluation of processor performance.
For instance, prior work~\cite{nanobench_github,ccbench_github,googlebm_github,desikan2001measuring,anglegog-bm} develops microbenchmarks to calibrate architectural simulators against RTL, which can be used to assess key processor modules such as branch predictors and execution units. 
However, in most cases, microbenchmarks fail to isolate performance differences to a single microarchitectural feature, as they are often affected by multiple interacting components.
Ideally, each benchmark should capture only one source of discrepancy in the simulator to facilitate targeted calibration.

\textbf{Time-proportional Method (e.g., TEA/TIP):}
Time-proportional analysis~\cite{gottschall2023tea,gottschall2021tip}, widely used in software profilers, attributes performance bottlenecks to individual instructions by ranking them according to the number of cycles they occupy the head of the Reorder Buffer (ROB).
However, although effective in pinpointing bottlenecks for specific microarchitectural features, these methods fail to expose microarchitectural discrepancies across different processors, since the resulting bottleneck rankings are not directly comparable.
This limitation stems from the fact that Time-proportional methods are primarily designed to analyze performance within a single microarchitecture to guide software optimizations, rather than to compare behaviors across different hardware configurations.

\textbf{Performance Event Analysis:}
Performance event analysis attributes performance bottlenecks to microarchitectural components~\cite{yasin2014top,eyerman2006performance}.
However, event counters are not necessarily directly correlated with performance impact~\cite{gottschall2023tea}.
Multiple correlated events may exhibit anomalies simultaneously, and domain expertise remains essential to distinguish relevant counters from noise to accurately identify the root cause.
Moreover, they have the same incomparability problem as the Time-proportional method, as shifts in bottlenecks do not directly imply the microarchitectural discrepancies between the two implementations.

Overall, existing performance analysis and calibration methods face challenges in single microarchitectural feature attribution, making calibration difficult.

\section{Motivation: Single Feature Attribution} \label{motivation}
Behavior-driven calibration is an indispensable step in calibrating the architectural simulator with the RTL.
Although we already have access to the RTL code and feature list, behavior-driven calibration enables us to effectively detect 
outdated parameter configurations, idealized or incomplete resource-contention modeling, or misestimated microarchitectural feature importance that may remain in specification-driven calibration.

\begin{table}[h!]
  \centering
  \caption{Top 10 bottleneck instructions identified by TEA on two hardware configurations. These results show a differing attribution order making root-cause analysis difficult.}
  \label{table:instr-tea}
  \renewcommand{\arraystretch}{1}
  \resizebox{0.45\textwidth}{!}{
  \begin{tabular}{
    c
    l
    c
    l
  }
    \toprule
    \multicolumn{2}{c}{\textbf{KMH-XS-RTL}} & 
    \multicolumn{2}{c}{\textbf{MINIMAL-XS-RTL}}\\
    
    PC & \multicolumn{1}{c}{Top 10 Instructions} & PC & \multicolumn{1}{c}{Top 10 Instructions}\\
    \midrule
    \textit{0x83f98} & \textit{fadd.d  fa0,fa0,ft1} & \textit{0x5aba8} &\textit{ld a5,0(a5)}\\
    \textit{0x5ab98} & \textit{ld a1,48(s1)}        & \textit{0x5ab98} &\textit{ld a1,48(s1)}\\
    \textit{0x26c70} & \textit{fmadd.d fa4,fa4,fa1,fa3} & \textit{0x417dc} &\textit{ld a5,8(a5)}\\
    \textit{0x26cfa} & \textit{fneg.d fa5,fa5} & \textit{0x5abe0} &\textit{lw a5,8(s0)}\\
    \textit{0x417e0} & \textit{snez a0,a0}    &  \textit{0x5ab94} &\textit{ld s1,16(s1)}\\
    \textit{0x226e6} & \textit{beqz a5,226bc} &  \textit{0x26cfa} &\textit{fneg.d fa5,fa5}\\
    \textit{0x226a6} & \textit{bnez a5,226c4} &  \textit{0x5abb0} &\textit{lw a5,20(s2)}\\
    \textit{0x5abe0} & \textit{lw a5,8(s0)}   &  \textit{0x5aba2} &\textit{ld a5,0(s1)}\\
    \textit{0x26d06} & \textit{beqz a5,26d82} &  \textit{0x5abf2} &\textit{ld s0,16(s0)}\\
    \textit{0x5a6bc} & \textit{lw a5,8(s0)}   &  0x\textit{}226c8 &\textit{fld fa1,-1944(gp)}\\
    \bottomrule
  \end{tabular}
  }
  \end{table}

Existing behavior-driven calibration approaches can pinpoint discrepancies between the simulator and RTL.
For example, TEA attributes performance differences to individual instructions.
However, these discrepancies often stem from interactions among multiple microarchitectural components and fail to provide fine-grained guidance for calibration.
We implement TEA~\cite{gottschall2023tea} on XS-RTL and evaluate it on two configurations: \textit{KMH}, a 6-wide high-performance core, and \textit{MINIMAL}, a simplified in-house design.
As shown in Table~\ref{table:instr-tea}, the bottleneck instructions identified in the two configurations differ significantly.
When different instructions are identified as bottlenecks across runs, the absence of a consistent baseline makes it difficult to provide accurate guidance for microarchitectural calibration. 
Moreover, bottleneck instructions are often associated with multiple microarchitectural features, further complicating the analysis.
While instruction-level attribution works well for analyzing a single processor, it faces challenges when applied to cross-architecture calibration.

To accurately identify the root causes, behavior-driven calibration approaches (e.g., microbenchmarks) should produce outputs that correlate closely with the targeted microarchitectural feature.
Although extensive efforts have been made to design targeted benchmarks for detecting functional and performance bugs or exploring simulator details~\cite{mammo2016bugmd,jo2018diagsim,singhal2004performance,yotov2005x,zen4-bm,BarbozaJKKGH21}, benchmarks aimed at verifying the consistency between architectural simulators and RTL implementations remain relatively underexplored.
We make the insight that \textit{a single benchmark is often insufficient to capture the entire state space of a given microarchitectural feature, and is prone to interference from unrelated factors}.

Based on these observations, we propose \textit{Microarchitecture Cliffs (Cliffs)}, a methodology that systematically constructs targeted \textit{Cliff benchmarks} to expose the behavior of a single microarchitectural feature, enabling effective diagnosis of microarchitectural feature-level inconsistencies.
The methodology begins by identifying key architectures that require calibration and generates benchmark sets for each critical microarchitectural feature to localize performance discrepancies between the simulator and RTL. 
In addition, we develop automated tools to enhance the efficiency of this process.

In particular, Cliff benchmarks use instruction snippets tailored to specific microarchitectural components to enable precise external observation of these features. 
The outputs of Cliff benchmarks produce a correlated mapping between the degree of stress applied and the microarchitectural response, typically reflected in metrics such as execution cycles or IPC. 
These performance trends, including non-linear responses identified by inflection points in the state space, provide valuable insights into the modeled behavior of the microarchitectural feature.

\begin{figure}[tbp]
\centering
\includegraphics[width=\linewidth]{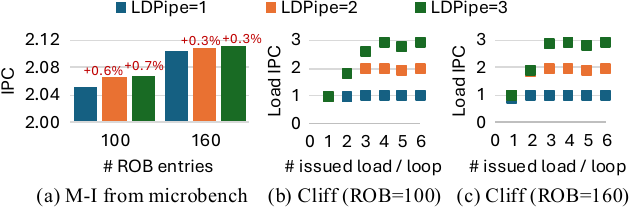}
\caption{
{M-I microbenchmark vs. Cliffs for L1 DCache bandwidth. M-I shows minimal IPC sensitivity to LDPipe, while Cliffs reveal clear distinctions among pipeline widths.}}
\label{fig:cliff-pk-mi}
\end{figure}

To illustrate the importance of Cliffs' single-feature attribution, we compare Cliff benchmarks against the M-I microbenchmark for independent memory accesses from a widely used calibration suite~\cite{desikan2001measuring,microbench_github}.
As shown in Fig.~\ref{fig:cliff-pk-mi}, both benchmarks aim to measure the effective bandwidth of the L1 DCache (which we call LDPipe).
However, M-I's testing effectiveness falls short of expectations:
increasing LDPipe from 1 to 3 raises the IPC by less than 1\% across different ROB configurations.
Consequently, using M-I under the assumption that LDPipe miscalibration will manifest as noticeable IPC discrepancies is ineffective, as the variation induced by LDPipe is even smaller than that caused by unrelated microarchitectural features such as ROB size.

In contrast, Cliffs clearly differentiates the behaviors of LDPipe=1, 2, and 3.
As the number of issued loads per loop increases, the load IPC saturates at distinct levels, forming well-separated plateaus corresponding to different LDPipe configurations.
This quantitative separation directly reflects the underlying bandwidth constraint and demonstrates Cliff's ability to attribute performance differences to the LDPipe feature alone, thereby enabling reliable feature-aware calibration.

As mentioned in $\S$\ref{b-forwardanalysis}, most high-level simulators do not implement every microarchitectural detail of the RTL design.
The Cliff methodology provides an effective way to assess how accurately a simulator reflects the underlying hardware. 
By operating at the microarchitectural feature level, it enables us to determine which components are faithfully modeled and which are insufficient, thereby revealing whether the simulator is suitable for evaluating a set of optimization ideas.

\section{Design} \label{design}

The design of the Cliff methodology faces two key technical challenges.
First, how to select the key architectures that require calibration; 
second, how to construct Cliff benchmarks for these features to enable single feature attribution.

\begin{figure}[tp]
\centering
\includegraphics[width=0.35\textwidth]{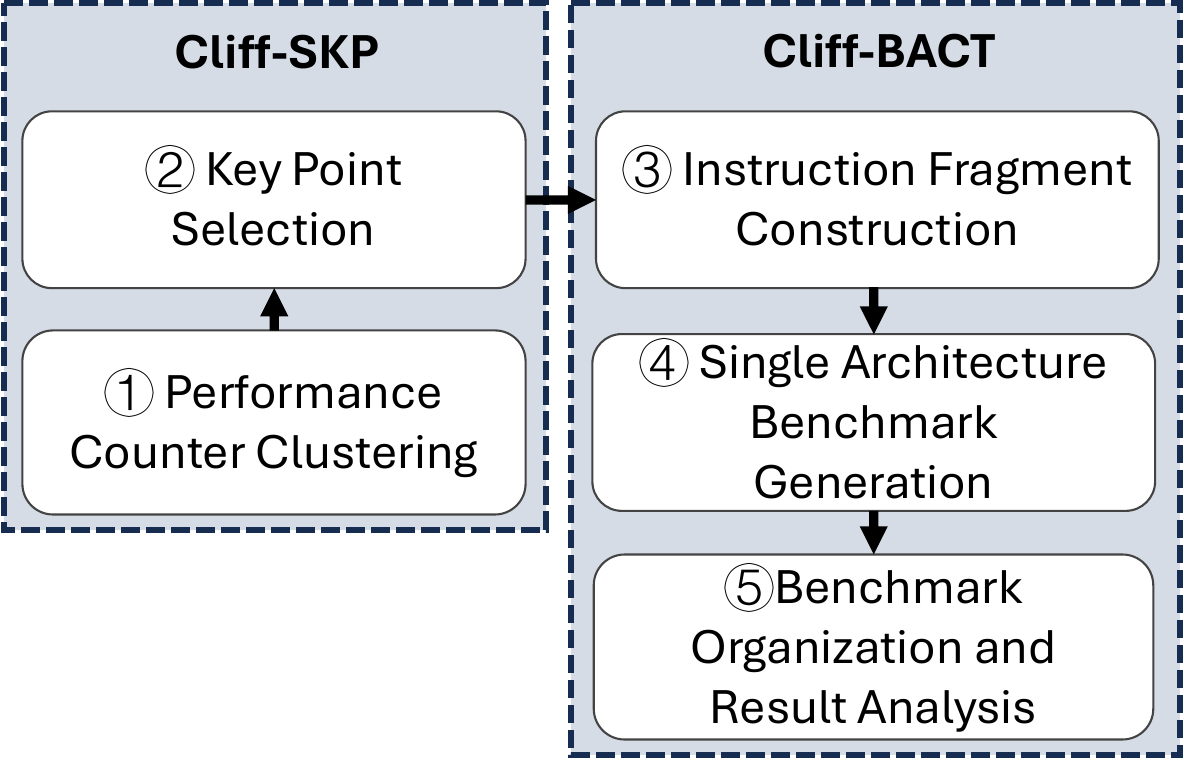}
\caption{Overview of the Cliff methodology. Cliff-SKP clusters performance-counter probes to identify key architectural bottlenecks. Based on the key architectures, Cliff-BACT constructs Cliff benchmarks that isolate individual microarchitectural traits for single-feature performance attribution}
\label{fig:cliffpip}
\end{figure}

To address these challenges, the Cliff methodology is designed with two main components, as illustrated in Fig.~\ref{fig:cliffpip}.
Cliff-SKP (Selecting Key architectural bottleneck Points) identifies performance counters that reflect architectural bottlenecks as \textit{information probes} and clusters these counters to pinpoint bottleneck architectures. 
Next, after identifying key architectures, Cliff-BACT (Benchmark Assembly for Characterizing Traits) extends these microarchitectural features to construct Cliff benchmarks, enabling single-feature attribution.

\subsection{Cliff-SKP}\label{sec:design_cliffchoose}

To prioritize the calibration of architectural components with significant performance impact, Cliff-SKP uses performance counters that reflect architectural bottlenecks as \textit{information probes}. 
It executes SPEC CPU2006 on XS-GEM5 during the initial calibration phase, 
followed by clustering on the collected performance counter data.
Based on the results, it identifies critical architectural regions as key directions for constructing Cliff benchmarks.
The clustering process consists of two rounds.
The first round broadly categorizes performance bottlenecks into either frontend-bound or backend-bound, while the second round further breaks down the specific architectural components that contribute most significantly to the relevant category.

The first-round clustering begins with the top-level performance counters defined by Top-Down~\cite{yasin2014top}, which organizes performance bottlenecks into four key categories: Frontend Bound, Bad Speculation, Backend Bound, and Retiring. 
These counters provide a coarse-grained foundation for identifying 
architectural bottlenecks.
We then apply the DBSCAN~\cite{ester1996density} clustering algorithm to analyze their behavior, and visualize the results with a heatmap. 
With the exception of a few outliers, most test points are clustered under the Backend Bound category.
This indicates that the main performance bottleneck lies in the processor backend, providing a clear direction for the design of Cliff benchmarks.

\begin{figure}[tbp]
\centering
\includegraphics[width=0.45\textwidth]{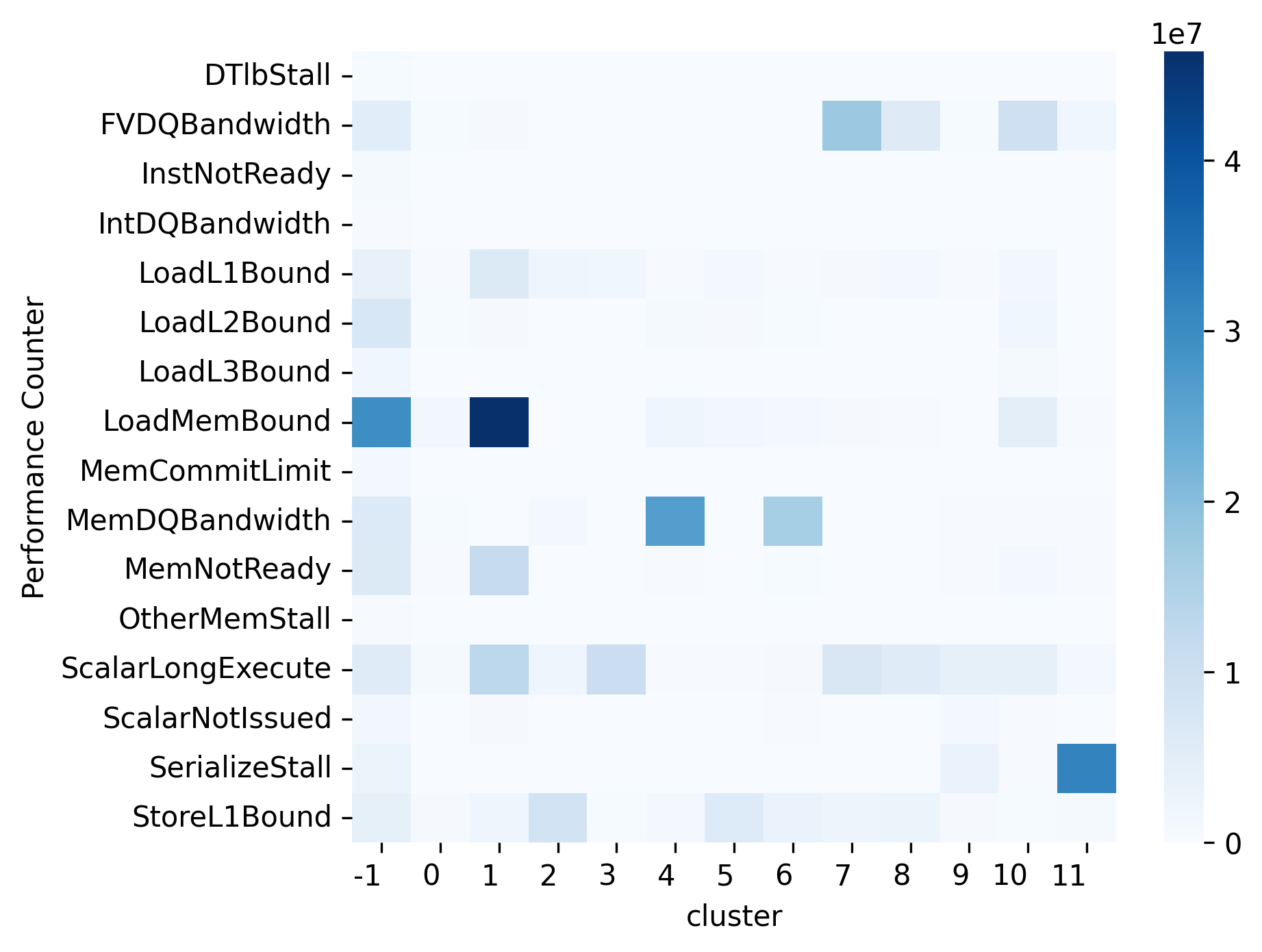}
\caption{Second-round clustering of backend performance counters identifies floating-point bandwidth, memory instruction bandwidth, L1/L2 caches, and system memory as concrete directions for Cliff benchmark construction.}
\label{fig:cluster2}
\end{figure}

Building on the initial clustering results, we conduct a finer-grained clustering analysis on the 16 performance counters categorized under backend bottlenecks in the Top-Down methodology, which further breaks down the causes of backend bottlenecks.
As shown in Fig. \ref{fig:cluster2}, the refined clustering allows us to preliminarily exclude architectural components with relatively minor impact on overall performance, such as the DTLB (DTlbStall), integer bound (IntDQBandwidth), L3 cache (LoadL3Bound), and commit limit (MemCommitLimit). 
Ultimately, we focus Cliff benchmarks construction on key backend architectures including floating-point bound (FVDQBandwidth), memory instruction (MemDQBound), L1 cache (LoadL1Bound), L2 cache (LoadL2Bound), and system memory (LoadMemBound), which are identified as the most likely to have the greatest performance impact.

\subsection{Cliff-BACT}\label{sec:cliff_design}

Given the identified target architectural components, we first construct Cliff benchmarks based on prioritized microarchitectural features, 
ensuring that each benchmark reflects a single microarchitectural feature.

Unlike traditional microbenchmarks that rely on a single program to evaluate performance, the Cliff methodology employs carefully designed groups of instruction snippets. 
By analyzing the IPC or execution time trends of these snippet groups, one can identify differences in the target microarchitectural feature between the simulator and RTL, and assess the degree of calibration achieved.
Cliff benchmarks start from parameter-level features and extend known behaviors using instruction snippet groups with performance trends, ensuring both visibility and isolation of the target feature.
The Cliff methodology constructs Cliff benchmarks that need to bridge the gap between instruction-level behavior and microarchitectural level, 
while reducing interference between different microarchitectural features within a single benchmark.
For complex structures, accurate evaluation may require combining multiple instruction snippets and incorporating observations from related components.

\subsubsection{Bridge the Gap Between the Instruction Level and the Microarchitecture Level}

To bridge the gap between the instruction level and the microarchitecture level, we construct instruction snippets that make specific microarchitectural features the performance bottlenecks. 
Benchmarks are categorized into functional classes, each with a tailored construction method, and microarchitectural pressure is progressively increased within each category.
By analyzing performance trends and identifying inflection points, we systematically reveal the features of targeted microarchitectural components.

\textit{Latency Testing:} By constructing read-after-write (RAW) dependency chains between instructions, the execution latency of the target instruction is the bottleneck on the critical path.

\textit{Bandwidth Testing:} Independent instruction streams without dependencies are constructed to expose execution bandwidth bottlenecks in target microarchitectural components.

\textit{Capacity Test:}
This benchmark combines long-latency and short-latency instructions to apply structural pressure.
A long-latency instruction blocks the release of resources, so the maximum capacity can be determined by measuring the execution time with different numbers of short-latency instructions.

\textit{Special Case Testing:} 
Certain microarchitectural features can be evaluated by designing paired benchmarks, where one benchmark triggers the feature and the other does not.
This approach enables researchers to determine whether the processor implements a specific optimization.

\subsubsection{Reduce Interference Between Different Microarchitectural Features Within a Single Benchmark}
To ensure accurate and valid measurement, effective Cliff benchmark design relies on \textit{interference suppression} and \textit{causal observability}.
The interference suppression principle requires minimizing the influence of irrelevant microarchitectural features during the execution of instruction snippets. 
To ensure that each Cliff benchmark reflects only the effect of the targeted microarchitectural feature, interference from other features on IPC should be minimized.
Additionally, Cliff benchmarks should warm up the branch predictor through repeated execution and constrain the memory footprint size to ensure that the target cache block resides in the intended cache level prior to measurement. 
These precautions help produce stable and reliable results.

The causal observability principle is grounded in the pressure gradient construction methodology. 
Each Cliff benchmark consists of a set of instruction snippets that reflect the characteristics of a single microarchitectural feature.
When the set of instruction snippets applies progressively increasing pressure to the target microarchitectural feature, its execution time or IPC should exhibit a corresponding regular variation.

\subsubsection{Organizing Cliff Benchmarks by Microarchitectural Feature}

Some microarchitectural features span multiple related components. 
Cliff benchmarks can start by using simple instruction snippets to test relatively independent baseline structures. 
Once the performance characteristics of these simpler components are well understood, more complex microarchitectural traits can be progressively identified. 
Subsequently, by systematically organizing and coordinating multiple Cliff benchmarks, the overall behavior of the target microarchitectural feature can be accurately captured.

\subsubsection{Identifying Undervalued Microarchitectural Features}
After constructing Cliff benchmarks based on prioritized microarchitectural features, we analyze abnormal trends or deviations from expected behavior 
to reveal microarchitectural features that might initially be deemed unimportant.
At the same time, this process serves as a \textit{microarchitectural audit} of the original design and specification-driven calibration flow, helping identify  limitations in the calibration process.

Next, we use two case studies to demonstrate (1) how to systematically construct Cliff benchmarks and (2) how we can discover unexpected discrepancies due to undervalued performance features.

 \subsection{Cliffs Design Case Study 1: ROB}\label{casestudy_rob}

This case study illustrates how to systematically construct Cliff benchmarks to reveal individual microarchitectural features, using our experience in building ROB-related Cliff benchmarks as an example.

\begin{figure}[htbp]
\centering
\includegraphics[width=0.48\textwidth]{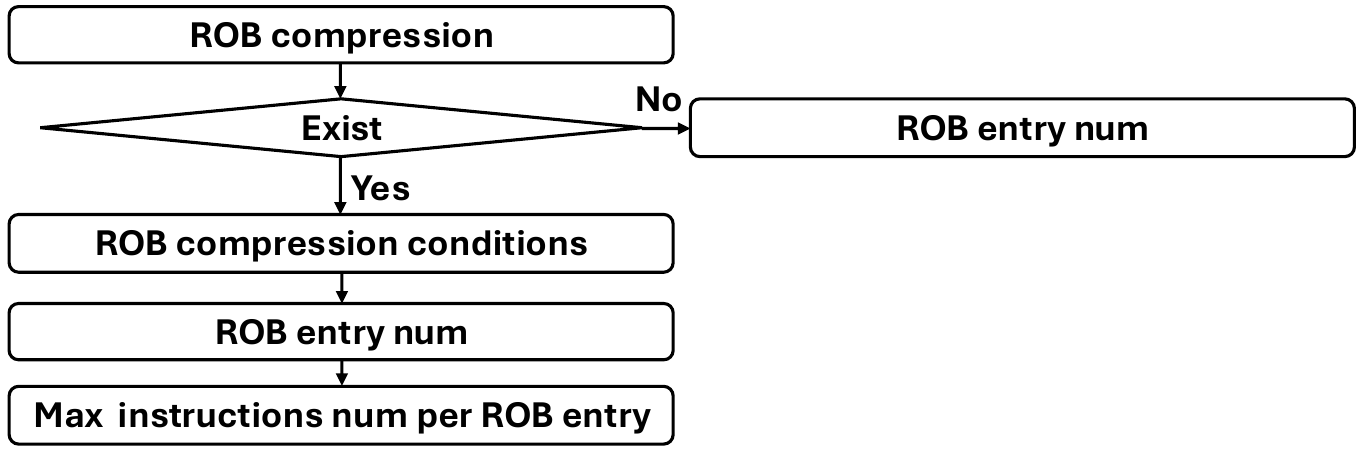}
\caption{ROB-related Cliff benchmarks.}
\label{fig:robcliffs}
\end{figure}

\textit{Complexities and solutions.}
ROB compression packs eligible instructions into a single ROB entry.
This significantly increases the effective capacity of the ROB, thereby enhancing overall processor performance.
To evaluate ROB-related features, we construct a series of Cliff benchmarks.
When the ROB can hold all instructions, the execution time remains constant due to parallelism; once its capacity is exceeded, the execution time increases.
As shown in Fig.~\ref{fig:robcliffs}, the overall evaluation procedure consists of the following three stages: (1) determining whether ROB compression exists; (2) identifying the conditions under which compression occurs and the compression capacity per entry; (3) measuring the maximum number of instructions a single ROB entry can compress.

\subsubsection{Determining the Existence of ROB Compression}

To verify the performance correctness of ROB compression feature modeling in the simulator, we construct Cliff benchmarks that use different combinations of instruction types occupying the ROB.
Therefore, we probe the ROB behavior using various permutations of six instruction types: \textit{beq}, \textit{ld}, \textit{st}, \textit{add}, \textit{fadd}, and \textit{nop}.
To bridge the gap between the instruction level and the microarchitectural level,
we adopt the approach described in the \textit{capacity test} to evaluate ROB size: 
constructing a long-latency load dependency chain to stall the head of the ROB, while filling the remaining entries with short-latency instructions selected from the six instruction types discussed above.
As the number of short-latency instructions increases, more ROB entries are occupied, leading to higher pressure on ROB. 
Once the ROB can no longer accommodate these instructions, the execution time of the instruction snippet starts to rise.
This marks the inflection point in the test results.

\begin{table}[htbp]
\caption{ %
Cliff benchmark results on calibrated XS-GEM5:
ROB capacity under different instruction snippets.
Instr. Num records the number of instructions at which the execution time starts to increase
Performance inflection points indicate bottlenecks related to ROB compression capacity. 
Background colors highlight regions where instruction snippets are influenced by other microarchitectural constraints.
}
\label{tbl:robrelatedresult}
\setlength{\tabcolsep}{0.28em}
\resizebox{\columnwidth}{!}{%
\begin{tabular}{rr|rcr|r|rcrcr|r|}
\hline
\multicolumn{1}{|r|}{\begin{tabular}[c]{@{}r@{}}Instr. \\  Fragments\end{tabular}} & \multicolumn{1}{r|}{\begin{tabular}[c]{@{}r@{}}Instr. \\  Num\end{tabular}} & \multicolumn{3}{r|}{\begin{tabular}[c]{@{}r@{}}Instr. \\  Fragments\end{tabular}}      & \begin{tabular}[c]{@{}r@{}}Instr. \\  Num\end{tabular} & \multicolumn{5}{r|}{\begin{tabular}[c]{@{}r@{}}Instr. \\  Fragments\end{tabular}}                                                                 & \multicolumn{1}{r|}{\begin{tabular}[c]{@{}r@{}}Instr. \\  Num\end{tabular}} \\ \hline
\multicolumn{1}{|r|}{\cellcolor[HTML]{FFF2CC}ld}                                   & \cellcolor[HTML]{FFF2CC}72                                                  & \cellcolor[HTML]{FFF2CC}ld  & \cellcolor[HTML]{FFF2CC}+ & \cellcolor[HTML]{FFF2CC}nop  & \cellcolor[HTML]{FFF2CC}144                            & beq                         & +                         & ld                           & +                         & st                           & 152                                                                         \\
\multicolumn{1}{|r|}{\cellcolor[HTML]{FCE4D6}st}                                   & \cellcolor[HTML]{FCE4D6}56                                                  & \cellcolor[HTML]{FFF2CC}ld  & \cellcolor[HTML]{FFF2CC}+ & \cellcolor[HTML]{FFF2CC}add  & \cellcolor[HTML]{FFF2CC}144                            & beq                         & +                         & ld                           & +                         & nop                          & 152                                                                         \\
\multicolumn{1}{|r|}{nop}                                                          & \multicolumn{1}{r|}{\textgreater 350}                                       & \cellcolor[HTML]{FFF2CC}ld  & \cellcolor[HTML]{FFF2CC}+ & \cellcolor[HTML]{FFF2CC}fadd & \cellcolor[HTML]{FFF2CC}144                            & beq                         & +                         & ld                           & +                         & add                          & 152                                                                         \\
\multicolumn{1}{|r|}{\cellcolor[HTML]{E2EFDA}fadd}                                 & \cellcolor[HTML]{E2EFDA}160                                                 & \cellcolor[HTML]{FFF2CC}ld  & \cellcolor[HTML]{FFF2CC}+ & \cellcolor[HTML]{FFF2CC}beq  & \cellcolor[HTML]{FFF2CC}144                            & beq                         & +                         & ld                           & +                         & fadd                         & 152                                                                         \\
\multicolumn{1}{|r|}{\cellcolor[HTML]{D9E1F2}add}                                  & \cellcolor[HTML]{D9E1F2}182                                                 & \cellcolor[HTML]{FCE4D6}st  & \cellcolor[HTML]{FCE4D6}+ & \cellcolor[HTML]{FCE4D6}nop  & \cellcolor[HTML]{FCE4D6}112                            & beq                         & +                         & st                           & +                         & nop                          & 152                                                                         \\
\multicolumn{1}{|r|}{beq}                                                          & 152                                                                         & \cellcolor[HTML]{FCE4D6}st  & \cellcolor[HTML]{FCE4D6}+ & \cellcolor[HTML]{FCE4D6}beq  & \cellcolor[HTML]{FCE4D6}112                            & beq                         & +                         & st                           & +                         & add                          & 152                                                                         \\ \cline{1-2}
\multicolumn{1}{l}{}                                                               &                                                                             & \cellcolor[HTML]{FCE4D6}st  & \cellcolor[HTML]{FCE4D6}+ & \cellcolor[HTML]{FCE4D6}ld   & \cellcolor[HTML]{FCE4D6}112                            & beq                         & +                         & st                           & +                         & fadd                         & 152                                                                         \\
\multicolumn{1}{l}{}                                                               &                                                                             & \cellcolor[HTML]{D9E1F2}nop & \cellcolor[HTML]{D9E1F2}+ & \cellcolor[HTML]{D9E1F2}add  & \cellcolor[HTML]{D9E1F2}380                            & \cellcolor[HTML]{D9E1F2}beq & \cellcolor[HTML]{D9E1F2}+ & \cellcolor[HTML]{D9E1F2}add  & \cellcolor[HTML]{D9E1F2}+ & \cellcolor[HTML]{D9E1F2}add  & \cellcolor[HTML]{D9E1F2}240                                                 \\ \cline{1-2}
\multicolumn{2}{|c|}{\textbf{Bottlenecks}}                                                                                                                       & \cellcolor[HTML]{E2EFDA}add & \cellcolor[HTML]{E2EFDA}+ & \cellcolor[HTML]{E2EFDA}fadd & \cellcolor[HTML]{E2EFDA}324                            & \cellcolor[HTML]{E2EFDA}beq & \cellcolor[HTML]{E2EFDA}+ & \cellcolor[HTML]{E2EFDA}fadd & \cellcolor[HTML]{E2EFDA}+ & \cellcolor[HTML]{E2EFDA}fadd & \cellcolor[HTML]{E2EFDA}240                                                 \\ \cline{1-2}
\multicolumn{2}{|c|}{\cellcolor[HTML]{FFF2CC}load queue size}                                                                                                    & beq                         & +                         & fadd                         & 152                                                    & beq                         & +                         & nop                          & +                         & nop                          & 320                                                                         \\ \cline{1-2}
\multicolumn{2}{|c|}{\cellcolor[HTML]{FCE4D6}store queue size}                                                                                                   & beq                         & +                         & nop                          & 152                                                    & \cellcolor[HTML]{D9E1F2}beq & \cellcolor[HTML]{D9E1F2}+ & \cellcolor[HTML]{D9E1F2}nop  & \cellcolor[HTML]{D9E1F2}+ & \cellcolor[HTML]{D9E1F2}add  & \cellcolor[HTML]{D9E1F2}180                                                 \\ \cline{1-2}
\multicolumn{2}{|c|}{\cellcolor[HTML]{D9E1F2}int reg num}                                                                                                        & beq                         & +                         & add                          & 152                                                    & \cellcolor[HTML]{E2EFDA}beq & \cellcolor[HTML]{E2EFDA}+ & \cellcolor[HTML]{E2EFDA}nop  & \cellcolor[HTML]{E2EFDA}+ & \cellcolor[HTML]{E2EFDA}fadd & \cellcolor[HTML]{E2EFDA}180                                                 \\ \cline{1-2}
\multicolumn{2}{|c|}{\cellcolor[HTML]{E2EFDA}fp reg num}                                                                                                         & \multicolumn{1}{r}{}        & \multicolumn{1}{r}{}      & \multicolumn{1}{r|}{}        & \multicolumn{1}{l|}{}                                  & \cellcolor[HTML]{E2EFDA}beq & \cellcolor[HTML]{E2EFDA}+ & \cellcolor[HTML]{E2EFDA}fadd & \cellcolor[HTML]{E2EFDA}+ & \cellcolor[HTML]{E2EFDA}add  & \cellcolor[HTML]{E2EFDA}180                                                 \\ \hline
\end{tabular}
}
\end{table}

To isolate the ROB as the performance bottleneck, we measure multiple combinations of instruction snippets.
A single \textit{ld} instruction is affected by both the load queue and the ROB. 
By gradually reducing the load queue pressure (from \textit{ld} to \textit{ld+st} and then to \textit{beq+ld+st}), the inflection point eventually becomes determined solely by the ROB, effectively eliminating the impact of the load queue.
Similarly, as shown in Table~\ref{tbl:robrelatedresult}, sequences involving \textit{beq} or combinations such as\textit{ beq + add/fadd/nop} and \textit{beq + ld/st + nop/add/fadd}, exhibit the same inflection point, suggesting that a standalone \textit{beq} instruction cannot be compressed.
Experimental results show that the ROB holds 152 entries under non-compression conditions, with an additional 8 entries required to accommodate long-latency load instructions. This yields a total ROB capacity of 160 entries.

\subsubsection{Determining the Conditions for Compression}
To eliminate interference from the Load/Store Queue and register file pressure, we first fill the ROB with 130 \textit{beq} instructions, then issue the target instructions and observe the ROB saturation point.
The results show that \textit{ld} and \textit{st} instructions saturate the ROB at around 22 entries, %
indicating that they are incompressible.
In contrast, \textit{nop}, \textit{add}, and \textit{fadd} instructions exceed this limit, confirming they are compressible.

\subsubsection{Determining the Capacity for Compression}
To eliminate register file pressure, we fill the first 130 ROB entries with \textit{beq} instructions and use the remaining capacity to evaluate compression behavior.
Unexpectedly, the results show an average of 5.11 instructions per ROB entry, which is inconsistent with expected ROB behavior.
Further analysis of the Cliff benchmarks and RTL source code reveals that ROB compression is only performed on instructions dispatched in the same cycle.
To address this, we initially attempt to use correctly predicted conditional branches to control the boundaries of the test instructions. 
However, since conditional branches typically rely on an \textit{add} instruction to control loop iteration, and this data dependency makes it difficult to ensure that both instructions enter the ROB in the same cycle, preventing precise control.
We therefore switch to using unconditional jump instructions (\textit{jal}) combined with \textit{nop} instructions to construct the test snippet.
We use one \textit{jal} followed by \textit{N} \textit{nops}, i.e., \textit{jal + N × nop}), where \textit{N} is the number of \textit{nops} in each instruction snippet.
Since \textit{jal} is not compressible but \textit{nops} are, each group of \textit{jalr} + \textit{nops} instructions occupies exactly two ROB entries when \textit{N} is smaller than the compression capacity.

Results show that when \textit{N} is less than or equal to 6, the ROB capacity is 64 groups of test snippets.
When \textit{N} exceeds 6, the effective ROB capacity starts to decrease, indicating that the maximum number of compressible instructions per ROB entry is 6.
Additionally, by comparing the \textit{jal + nops} structure with snippets using only \textit{nop} instructions, we further confirm that only instructions fetched in the same cycle can be compressed into a single ROB entry,
as successful jumps break the fetch continuity across cycles.

\subsection{Cliffs Design Case Study 2: DCache Banks}
\label{sec:casestudy-bank}
\begin{figure}[htbp]
\centering
\includegraphics[width=0.48\textwidth]{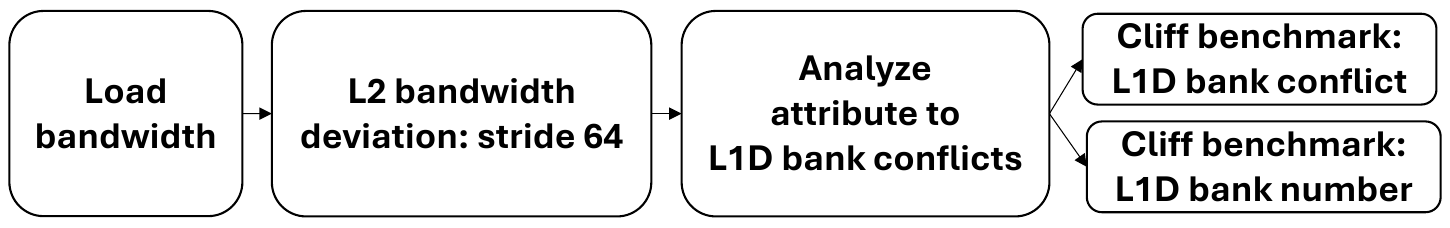}
\caption{Analyzing performance deviations.}
\label{fig:foundcliff}
\end{figure}

The Cliff methodology focuses on single-feature performance validation.
When constructing Cliff benchmarks, we use our domain knowledge of the target architecture (XS-RTL) to obtain a comprehensive list of features to be tested and calibrated.
This design is a necessary step, as specification-driven calibration from design specifications may introduce discrepancies that are difficult to capture statically through code reviews.
This case study illustrates how analyzing performance deviations can reveal microarchitectural features that have been undervalued for its performance impact initially, thereby motivating the construction of additional Cliff benchmarks. 
The overall workflow is summarized in Fig.~\ref{fig:foundcliff}.

As discussed in $\S$\ref{sec:design_cliffchoose}, clustering results from \textit{Cliff-SKP} highlight memory instructions and the L1 cache as primary targets for calibration. 
Accordingly, we construct Cliff benchmarks starting from the architectural parameters of these components.
With the load pipeline bandwidth already characterized, we proceed to evaluate L2 bandwidth under different load stride patterns. 
Specifically, RTL shows a significant drop in L2 bandwidth at a stride of 64 bytes, while XS-GEM5 does not. 
Based on the observed performance trends and the feature list, we identified that the discrepancies are primarily caused by L1D bank conflicts, a constraint that was deemed a low priority feature.
After confirming the phenomenon, we construct new Cliff benchmarks by varying the number of banks to characterize its impact, enabling fine-grained calibration between the simulator and RTL.

\section{Evaluation} \label{evaluation}

We evaluate the Cliff methodology from two major aspects. 
In the first part, we focus on evaluating Cliffs’ effectiveness.
We begin by examining the aggregate results of Cliffs ($\S$\ref{sec:cliff_results}), then assess its ability to reduce the relative error of feature evaluation ($\S$\ref{storeset}). We further evaluate its impact on lowering the simulator’s overall absolute performance error ($\S$\ref{overall-result}). 
Finally, we analyze the effect of the clustering process ($\S$\ref{cluster-result}) and present detailed results for selected Cliff benchmarks ($\S$\ref{cliff-result}).
In the second part, we examine whether Cliffs can accurately capture specific microarchitectural features of BOOM~\cite{celio2015berkeley} ($\S$\ref{boom-evaluation}).
We further adapt the methodology to new workloads beyond SPEC 
to demonstrate that Cliffs remains effective across different workloads ($\S$\ref{verilator}).

\subsection{Cliff Results}\label{sec:cliff_results}
\begin{figure}[htbp]
\centering
\includegraphics[width=0.48\textwidth]{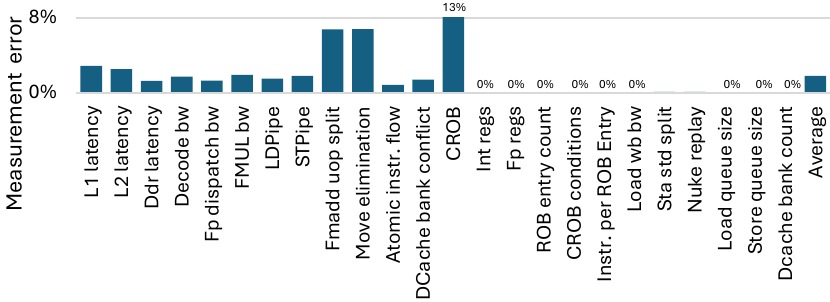}
\caption{Accuracy measurement of Cliff benchmarks shows an average error of only 1.8\% compared to the design values.}
\label{fig:overallaccuracy}
\end{figure}

By leveraging Cliff benchmarks, the Cliff methodology enables accurate and detailed characterization of microarchitectural features.
As shown in Fig.~\ref{fig:overallaccuracy}, Cliff benchmarks achieve a low overall deviation between the measured values and the design-intended values (measurement error) of just 1.8\%, demonstrating their strong evaluation capability.

\begin{table}[!ht]
  \centering
  \caption{Evaluation of microarchitectural features using Cliff benchmarks shows that calibration reduces the error on Cliff benchmarks by 57.8\%.} 
  \label{table:cliffresult}
  \renewcommand{\arraystretch}{1}
  \resizebox{\columnwidth}{!}{%
  \begin{tabular}{l S[table-format=3.2, table-number-alignment=right] S[table-format=3.2, table-number-alignment=right]@{\,}r@{\,\%)} S[table-format=3.2, table-number-alignment=right]@{\,}r@{\,\%)}}
    \toprule
    \multicolumn{1}{c}{\textbf{Test}} & \textbf{RTL} &
    \multicolumn{2}{c}{\textbf{Before}} &
    \multicolumn{2}{c}{\textbf{After}} \\
    \midrule
    L1 latency        & 4.07  & 4.14  & (+1.72  & 4.14  & (+1.72 \\
    L2 latency        & 16.07 & 27.08 & (+68.51 & 15.08 & (-6.16 \\
    Ddr latency       & 226.49 & 248.02 & (+9.51 & 218.48 & (-3.54 \\
    Decode bw         & 5.92  & 5.88  & (-0.68  & 5.88  & (-0.68 \\   
    LDPipe          & 2.96      & 2.95   & (-0.34  & 2.96   & (+0.00 \\
    STPipe         & 1.95      & 1.97   & (+1.03  & 1.97   & (+1.03 \\
    Int regs       &  224       &  224  & (+0.00 & 224   & (+0.00 \\
    Fg regs       &  192       &  192  & (+0.00 &  192   & (+0.00 \\
    ROB entry count       &  160       &  320  & (+100 &  160   & (+0.00 \\
    Load queue size      & 72        & 128    & (+77.78 & 72     & (+0.00 \\
    Store queue size     & 56       & 96     & (+71.43 & 56     & (+0.00 \\
    Fmadd uop split      & 4.10    & 2.18 & (-46.78 &  4.16 & (+1.61 \\
    Move elimination    &  5.65    & 5.56 & (-1.60  & 5.52 & (-2.30 \\
    Atomic instr. flow    &  33.19    & 13.16 & (-60.35  & 32.71 & (-1.45 \\
    Fp dispatch bw    & 2.97  & 3.94  & (+32.66 & 2.85  & (-4.04 \\
    FMUL bw           & 2.96  & 3.91  & (+32.09 & 2.95  & (-0.34 \\
    DCache bank conflict    &  0.99    & 2.95 & (+198  & 0.93 & (-6.06 \\
    CROB       &     4.60    &  1.00  & (-78.26 &  5.11   & (+11.09 \\
    CROB  conditions     & 4        & 0   & (-99.99 & 4    & (+0.00 \\
    Instr. per ROB Entry     &  6       & 1 & (-83.33 &  6   & (+0.00 \\
    Load wb bw     &  3       &  8  & (+166 &  3   & (+0.00 \\
    Sta std split        &  0       &  501  & (+99.99 &  0   & (+0.00 \\
    Nuke replay          &  0       & 501  & (+99.99 &  0   & (+0.00 \\

    DCache bank count         & 8         & 0      & (-99.99 & 8      & (+0.00 \\
    \bottomrule
  \end{tabular}
  }
\end{table}
Table~\ref{table:cliffresult} summarizes the results of Cliff benchmarks across various scenarios and reports their relative deviations from RTL.
Using Cliffs, we calibrate a set of microarchitectural features that are identified during specification-driven calibration.
Overall, the performance discrepancy between XS-GEM5 and XS-RTL on Cliff benchmarks is 59.2\% before calibration, and is reduced to 1.4\% after calibration. 
It is worth noting that we also evaluate the calibration of several detailed microarchitectural features, such as Nuke Replay, sta std split, and DCache bank conflict. After calibration, XS-GEM5 accurately models the corresponding pipeline behaviors observed in RTL.

\subsection{Relative Error of Store Set Evaluation}
\label{storeset}
To better evaluate Cliffs’ ability to guide simulator calibration and reduce relative error, we compare the Store Set evaluation on XS-RTL and XS-GEM5 before and after calibration. 
It is important to note that simulator mismatches do not only originate from incorrect parameter settings; they often stem from missing or incomplete microarchitectural features.

Before calibration, we observed significant discrepancies between XS-GEM5 and XS-RTL in the Store Set performance gains on \textit{h264ref}, with a relative error of 48.86\%.
However, solely observing the uncalibrated XS-GEM5 and XS-RTL behaviors on \textit{h264ref} makes it difficult to determine which microarchitectural features are responsible for these discrepancies.
As shown in Table~\ref{table:cliffresult}, the Cliff methodology reveals that two microarchitectural features related to memory-ordering dependencies (Nuke Replay and STA-STD separation) differ between the uncalibrated XS-GEM5 and XS-RTL.
\begin{figure}[tbp]
\centering
\includegraphics[width=\linewidth]{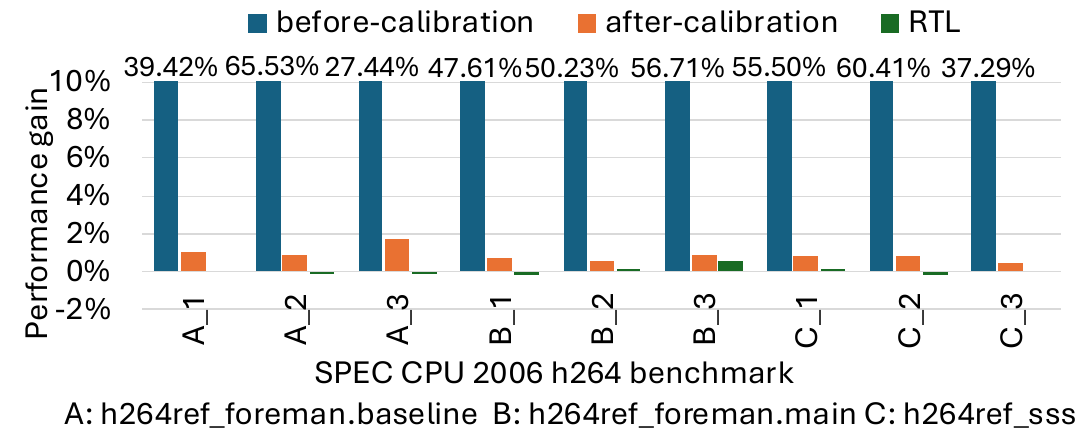}
\caption{Comparison of Store Set evaluation on \textit{h264ref} before and after calibration. The sensitivity is significantly larger on the uncalibrated simulator, illustrating a false trend.}
\label{fig:storeset}
\end{figure}

\begin{figure*}[!tbp]
\centering
\includegraphics[width=\textwidth]{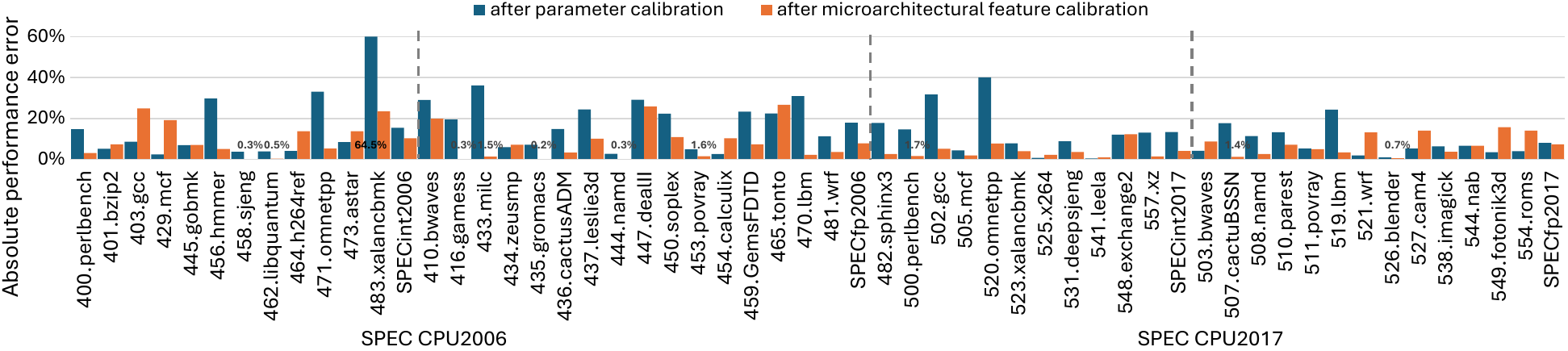}
\caption{%
Performance errors between XS-RTL and XS-GEM5 after parameter calibration and after microarchitectural feature calibration. SPECint or SPECfp errors are reported as average absolute errors across benchmarks.}
\label{fig:allspec}
\end{figure*}

With these microarchitectural features fixed, the simulator’s relative error in evaluating Store Set is significantly reduced to 0.83\%.
As shown in Fig.~\ref{fig:storeset}, before calibration, the simulator reports an average performance improvement of 48.90\% with Store Set on \textit{h264ref}, with the highest gain reaching 65.53\% for certain slices. 
As a result, the average performance benefit of Store Set drops to 0.87\%, and previously high-gain checkpoints show only 0.89\% improvement, which closely matches with the RTL behavior (0.04\% error).

These results demonstrate that, with Cliffs-assisted calibration, the architectural simulator can more accurately evaluate the true performance benefits of optimization techniques, effectively reducing the relative error.
Meanwhile, Cliffs help clarify which components of the simulator are faithfully modeled and which are insufficient, thereby enabling us to determine whether the simulator is suitable for evaluating this class of microarchitectural features.
In the Store Set evaluation, uncalibrated XS-GEM5 misrepresents strongly related features, resulting in high relative error, whereas calibrated XS-GEM5 correctly models these features and enables accurate evaluation of Store Set gains.

\subsection{Absolute Error} \label{overall-result}

We use absolute performance error to evaluate the overall calibration between the simulator and RTL.
To precisely assess the performance impact of Cliffs calibration on both parameter-level and microarchitecture-level features, we compare the performance errors before calibration, after parameter calibration, and after microarchitectural feature calibration.

As shown in Fig.~\ref{fig:allspec}, for SPECint2006, the absolute performance error between XS-GEM5 and XS-RTL decreases from 18.1\% before calibration to 15.6\% after parameter calibration, representing a 2.5\% reduction and indicating a substantial improvement in accuracy.
After calibrating the microarchitectural features, the absolute performance error further drops to 10.4\%, 
corresponding to a total reduction of 7.6\% compared with the before-calibration result and an additional 5.2\% improvement over the parameter-calibrated stage.
Similarly, for SPECfp2006, the absolute performance error is 23.1\% before calibration.
After parameter calibration, the error decreases to 18.0\%, representing a reduction of 5.1\%.
With further microarchitectural-feature calibration, the error drops to 8.0\%, corresponding to a total reduction of 15.1\% compared with the before-calibration result and an additional 10.1\% improvement over the parameter-calibrated stage.
For SPECint2017, parameter calibration reduces the absolute performance error from 19.3\% to 13.5\%, achieving a 5.8\% improvement.
With further microarchitectural-feature calibration, the error decreases to 4.2\%, representing a total reduction of 15.1\% compared with the before-calibration result and an additional 9.3\% improvement over the parameter-calibrated stage.
For SPECfp2017, the parameter calibration gap narrows from 28.5\% to 8.2\%, yielding a 20.3\% improvement.
With microarchitectural-feature calibration, the error decreases to 7.5\%, representing a total reduction of 21.0\% compared with the before-calibration result and an additional 0.7\% improvement over the parameter-calibrated stage.

Performance differences across some benchmarks exhibit local fluctuations, which is expected, as calibration does not reduce the gap in a strictly monotonic fashion. 
Nevertheless, the overall trend is clearly moving downward~\cite{black1998calibration}. 
The Cliff methodology effectively narrows the performance gap between XS-GEM5 and XS-RTL. 
Before calibration, opposing errors cancel out, concealing the true extent of model inaccuracies.
Partial calibration breaks this balance, revealing hidden inaccuracies and occasionally creating the illusion of reduced accuracy, while actually improving model fidelity.
For instance, although \textit{h264ref} appears less accurate after calibration, ($\S$\ref{storeset}) shows that the post-calibration trend better reflects RTL performance, enabling more faithful evaluation of Store Set.
These findings confirm that Cliffs enhance calibration accuracy.

\subsection{Performance Impact of Key Microarchitectural Bottlenecks}
\label{cluster-result}
Based on two rounds of performance counter clustering, we first identify the backend as the primary focus for constructing Cliff benchmarks. 
We then further refine the construction directions within the backend domain.

\begin{table}[tbp]
  \centering
  \caption{Performance counter metrics from Cluster 2-0.}
  \label{table:formatting2}
  \begin{tabular}{c l l}
    \hline
    \textbf{cluster} & \multicolumn{1}{c}{\textbf{key point}} & \multicolumn{1}{c}{\textbf{checkpoint}}\\
    \hline
        \multirow{3}{*}{2-0} & \multirow{3}{*}{\shortstack[l]{LoadMemBound\\ScalarLongExecute\\StoreL1Bound}}      & calculix\_1, \\ 
            &  & h264ref\_sss\_1,\\
            &       & xalancbmk\_1, wrf\_1\\
    \hline
  \end{tabular}
\end{table}

\begin{figure}[tbp]
\centering
\includegraphics[width=0.45\textwidth]{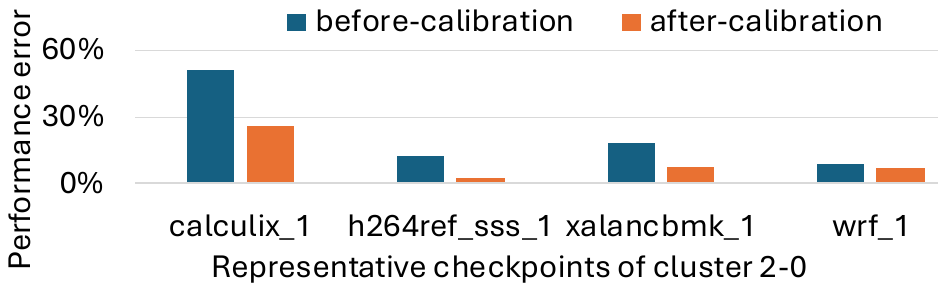}
\caption{Performance errors before and after calibration (Cluster2-0). Calibrating the critical performance factors correlates with reduction in performance gap on benchmarks belonging to the same cluster.}
\label{fig:clustercpt}
\end{figure}

Table~\ref{table:formatting2} presents the details of Cluster 2-0.
To further evaluate the impact of the clustered key factors on code segments, we analyze the effect of calibrating the critical performance features identified in Cluster 2-0. 
As shown in Fig.~\ref{fig:clustercpt}, after calibration, the performance gap between XS-GEM5 and XS-RTL on the key slices was significantly reduced, with the maximum reduction reaching 25.5\%, 
bringing the absolute error down to 1.2\%.

\subsection{Analysis of Cliff Benchmarks' Results}
\label{cliff-result}
As presented in ($\S$\ref{overall-result}), we demonstrate the overall evaluation effectiveness of the Cliff benchmarks. 
To further present their evaluation process in a more structured and detailed manner, we categorize the results into three types:
(1) trendline-based analyses for aiding result interpretation; (2) microarchitectural feature extraction through pairwise comparisons; (3) critical analysis based on identifying inflection points arising from variations in input parameters.

\subsubsection{Trend Line Guided Result Evaluation}
\begin{figure}[htbp]
\centering
\includegraphics[width=0.48\textwidth]{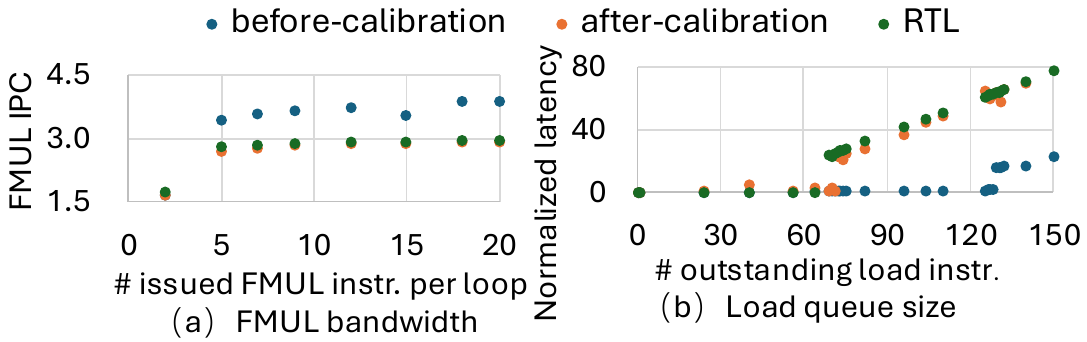}
\caption{Cliff benchmarks for FMUL bandwidth and load queue size. The performance profile of the calibrated model aligns with RTL.}
\label{fig:fmul-e}
\end{figure}

We employ trendline analysis to assist in evaluating the bandwidth of the floating-point multiplication (FMUL) unit. 
The benchmark results are shown in Fig.~\ref{fig:fmul-e}a. 
In this benchmark, we gradually increase the number of FMUL instructions that can be issued per cycle to incrementally stress the FMUL unit, while monitoring the corresponding changes in the instructions per cycle (IPC) for these operations. 
The results show that as the number of issued instructions increases, the IPC for FMUL instructions increases monotonically at first and then plateaus after reaching a peak value. 
This saturation point indicates the maximum bandwidth of the FMUL unit.

\subsubsection{Change Point Detection}
We evaluate the capacity of the load queue using a change point analysis approach. 
In our experiment, we construct an instruction snippet consisting of three dependent high-latency load operations, which serve to occupy and block several entries in the load queue. 
Concurrently, we gradually increase the number of parallel short-latency loads (i.e., outstanding loads) to continuously apply pressure on the load queue. 
As shown in Fig.~\ref{fig:fmul-e}b, by measuring the execution latency of the entire instruction segment, we observe a distinct latency jump when the number of outstanding loads reaches a certain threshold. 
This sudden increase in latency indicates the saturation point of the load queue, from which we can infer its upper-bound capacity.

\subsubsection{Architectural Characterization via Pairwise Comparison}
\begin{figure}[tbp]
\centering
\includegraphics[width=0.4\textwidth]{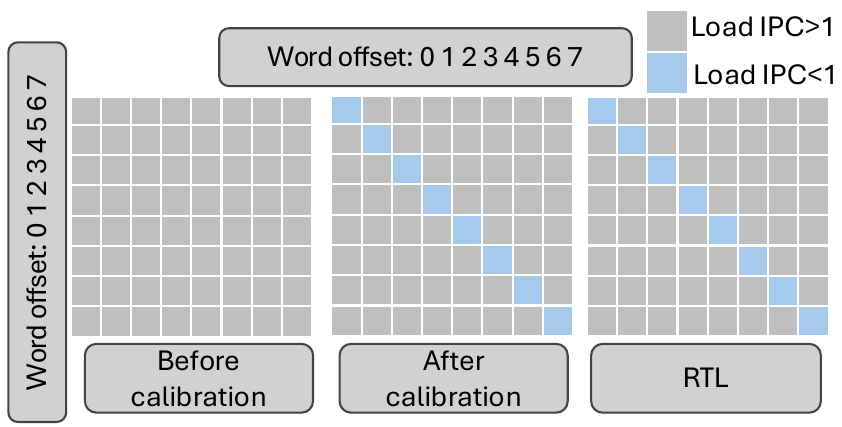}
\caption{Cliff benchmark for DCache banks. After calibration, the Cliff benchmark for DCache banks achieves performance behavior that closely matches the RTL results. }
\label{fig:l1dbanknum}
\end{figure}

To analyze the bank structure of the DCache, we focus on comparing the performance difference of load instructions when accessing the same byte versus different bytes. 
By constructing instruction sequences with varying memory offsets, we measure the IPC under both access patterns, as shown in Fig.~\ref{fig:l1dbanknum}. 
The results indicate that accessing the same word yields significantly lower IPC compared to accessing different words, suggesting the presence of bank conflicts that limit concurrent execution. 
This notable contrast implies that each word may map to a separate bank. Based on this observation, we infer that the DCache consists of 8 independent banks.

All three cases can accurately capture the corresponding microarchitectural features. After calibration, they can measure performance changes, accurately reflect architectural details, and quantitatively represent a single microarchitectural feature.



\subsection{Evaluating New Processors: BOOM}
\label{boom-evaluation}
\begin{figure}[tbp]
\centering
\includegraphics[width=0.45\textwidth]{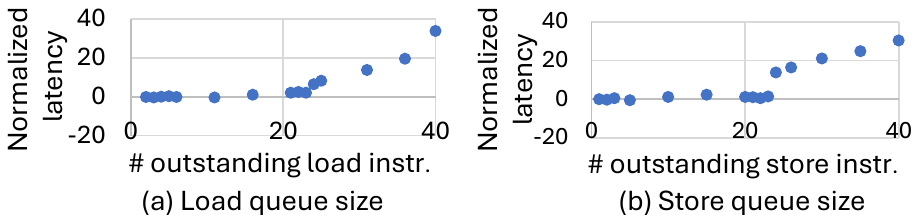}
\caption{Cliff benchmark for load/store queue size on BOOM.}
\label{fig:boom-e}
\vspace{-0.5cm}
\end{figure}

To evaluate whether the Cliff methodology can effectively capture architectural characteristics across different processors, we apply it to BOOM~\cite{celio2015berkeley} by constructing Cliff benchmarks targeting load queue size and store queue size. 
Fig.~\ref{fig:boom-e} illustrates the performance trend lines for these two benchmarks. 
A clear performance  inflection point is observed when the number of outstanding memory instructions reaches 24, resulting in a sharp increase in execution latency. 
This behavior precisely corresponds to BOOM’s microarchitectural constraint, where both the load and store queues are 24 entries in size.

These results demonstrate the generality and effectiveness of the Microarchitecture Cliffs methodology. 
Not only can it capture fine-grained differences in XS-GEM5 before and after calibration, as well as microarchitectural features of XS-RTL, but it also successfully reflects architectural characteristics in a significantly different microarchitecture like BOOM.

\subsection{Evaluating New Workloads: Verilator}
\label{verilator}
To better demonstrate the generality of the Cliff methodology across different scenarios, we use the \textit{Verilator} workload outside of the SPEC benchmark suite to evaluate the calibration effectiveness.
From the Cliff-SKP analysis, the clustering results indicate that the primary performance bottleneck of the Verilator workload lies in the frontend.
To address this bottleneck, we construct a Cliff benchmark to probe the maximum length of the Path History Register (PHR).
Specifically, we build two correlated conditional branches and progressively vary the number of intermediate branch instructions that occupy PHR bits between the correlated branches. %
When the PHR length becomes insufficient, the history of the first branch will be pushed out of the PHR, and the second branch exhibits a 50\% misprediction probability, leading to a sudden rise 
in the number of mispredicted branches.

\begin{figure}[tbp]
\centering
\includegraphics[width=0.48\textwidth]{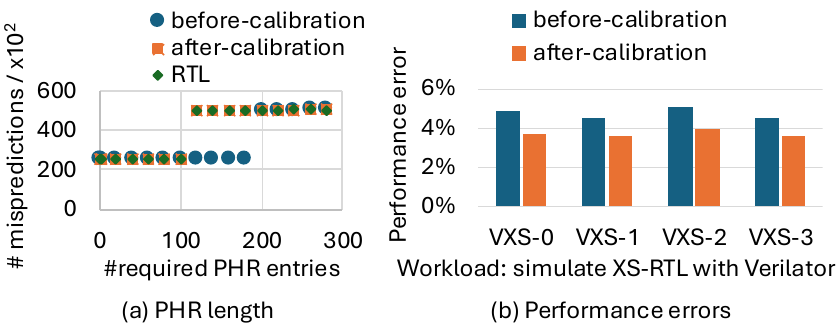}
\caption{
Cliff benchmark for PHR length and performance error 
running simulate XS-RTL with Verilator as a workload (VXS).
A spike in mispredictions indicates that the required PHR entries exceed the current length. 
Calibrating the PHR length effectively reduces frontend performance error.
}

\label{fig:veri}
\end{figure}

As shown in Fig.~\ref{fig:veri}, the maximum PHR length that sustains a low branch misprediction rate is approximately 200 in XS-GEM5, but significantly shorter, around 120, in XS-RTL.
This difference reveals a mismatch in the maximum PHR length between the two platforms. 
After calibrating the relevant architectural parameters, the performance trends of XS-GEM5 and XS-RTL converge, with both exhibiting a sharp rise of mispredictions at a history length of approximately 120.

To assess the impact of calibration, we compare the frontend performance error between XS-GEM5 and XS-RTL before and after calibration using the \textit{Verilator} workload.
The results show that after calibrating the PHR length, the frontend performance error of XS-GEM5 decreases significantly, dropping from 4.72\% to 3.73\%.

\section{Related Work} \label{related_work}

\textbf{Microbenchmarks for the Memory Subsystem:} These benchmarks primarily target the memory subsystem, highlighting critical metrics such as latency and bandwidth, with some 
extending to GPU memory~\cite{mei2016dissecting,fang2018benchmarking}. Nonetheless, attributing performance differences of instruction snippets to a single architectural factor remains a significant challenge~\cite{esmaili2024mess,fang2015measuring,verdejo2017microbenchmarks,verdejo2018main,mccalpin1995memory,mccalpin2006stream,ahmed2019hopscotch,staelin2005lmbench,hristea1997measuring,gomez2021benchmarking}.

\textbf{GPU Performance Calibration and Analysis Methodologies:}
Current GPU architecture analysis approaches mainly rely on the development and implementation of architecture-customized microbenchmarks. These testing tools primarily serve dual objectives: (1) evaluating the optimizability of GPU code through architecture-driven tests (such as analyzing memory hierarchy access patterns and stress testing computation unit throughput) to maximize computational throughput~\cite{yi2021cudamicrobench,fang2018benchmarking}; (2) designing workload-specific tests based on workload characteristics or hardware features to identify performance bottlenecks during software execution~\cite{taylor2010micro,khairy2020accel,ciznicki2012benchmarking,augonnet2010automatic,coleman2022wfbench}. 
However, the analysis results are still confined to software execution metrics,
without establishing a quantifiable architecture-performance relationship.

\textbf{Automatic Benchmark Generation: }
Automatic benchmark generation can create benchmarks based on the workload characteristics of applications, 
significantly reducing execution time while accurately reflecting the application's performance. 
This approach aims to minimize simulation time while accurately capturing the characteristics of the benchmarks. 
However, it still primarily focuses on software performance and results, with relatively little attention paid to hardware architecture features~\cite{bell2005improved,joshi2008automated,joshi2008return,joshi2007constructing,joshi2008distilling}.

Compared with memory subsystem microbenchmarks, GPU performance analysis methods, and automatic benchmark-generation,
the Cliffs attributes performance differences to single microarchitectural features and provides a more targeted link between microarchitecture and performance.

\section{Conclusion} ~\label{conclusion}
In this work, we introduce Microarchitecture Cliffs, a methodology for constructing benchmarks that isolate and attribute performance differences to single microarchitectural features.
Comprehensive evaluation shows that Cliff benchmarking narrows the performance gap between XS-GEM5 and XS-RTL, reducing the absolute performance error 
by 15.1\% on SPECint 2017 and 21.0\% on SPECfp 2017.
Moreover, in evaluating the representative Store Set algorithm, the calibrated simulator more faithfully reproduced RTL behavior and substantially reduced the relative error of the feature evaluation, further validating the effectiveness of Microarchitecture Cliffs in architectural calibration.

\bibliographystyle{ACM-Reference-Format}
\bibliography{refs.bib}

\end{document}